\documentclass[12pt]{iopart}
\begin{document}

\title[
Stationary quantum Markov process for the Wigner function
on a lattice phase space
]
{
Stationary quantum Markov process \\
\hspace{5mm} for the Wigner function on a lattice phase space
}

\author{
T.Hashimoto,
M.Horibe\(^*\),
A.Hayashi\(^\dagger\)
}

\address{
Department of Applied Physics,
University of Fukui \\
\hspace{10mm}
Bunkyo 3-9-1, Fukui 910-8507, JAPAN
}
\ead{
d901005@icpc00.icpc.fukui-u.ac.jp \\
\hspace{9.5mm}
\(*\)) horibe@quantum.apphy.fukui-u.ac.jp \\
\hspace{9.5mm}
\(\dagger\)) hayashi@soliton.apphy.fukui-u.ac.jp
}

\begin{abstract}
As a stochastic model for quantum mechanics
we present a stationary quantum Markov process
for the time evolution of the Wigner function
on a lattice phase space \(Z_N\times Z_N\) with \(N\) odd.
By introducing a phase factor extension to the phase space,
each particle can be treated independently.
This is an improvement on earlier methods
that require the whole distribution function
to determine the evolution of a constituent particle.
The process has branching and vanishing points, though
a finite time interval can be maintained between the branchings.
The procedure to perform a simulation using the process is presented.

\end{abstract}

\maketitle

\section{Introduction}
Various studies have been made of quantum mechanics
by using trajectories.
For example, in Feynman's path integral method,
the transition amplitude between the initial and final points
is calculated by adding all the phase factors determined from the trajectories
connecting these points [1].
In this case, the trajectories are interfering alternatives
and not those of observed particles.
The method proposed by Bohm is another example.
In this method, the trajectories are treated as if they were observed
and are calculated deterministically [2].
These methods give alternative formulations of
quantum mechanics.

The stochastic approach is another method.
It is used not only for the
formulation of quantum mechanics, but also for practical
calculations.
The use of a stochastic method for practical calculations is generally called the quantum Monte Carlo
method (QMC) and is widely used in the simulation of quantum systems.
Typical algorithms in the QMC are
the random walk Monte Carlo method (RWMC),
the diffusion Monte Carlo method (DMC),
and the path integral Monte Carlo method (PIMC).
These methods are efficient ways of determining
the wave function and energy of a ground state [3,4].
They are based on the formal analogy that
the Schr\"odinger equation becomes the diffusion equation
when \(t\) is substituted by an imaginary time \(i\tau\).

A solution of the diffusion equation can be constructed
using Brownian motion.
We can generate sample paths of Brownian particles
by a random number generator and can follow the paths
independently as Brownian motion is a stationary
Markov process,
which means that the diffusion equation can be solved
by a stochastic method using a random number generator.
Many studies have been conducted on the diffusion 
equation and Brownian motion.
It is known that a stochastic estimation is generally more
efficient than direct calculations when the system has many
degrees of freedom.  In direct methods, the amount of calculation
required generally increases exponentially with the number of
degrees of freedom \(D\) [5].
On the other hand, for stochastic methods the amount of calculation
is of order \(D\) for one trial, though the estimated error
is of order \(1/\sqrt{n}\) where \(n\) is the number of trials.
Very naively, the Schr\"odinger equation is a diffusion
equation along an imaginary time axis.
However, the quantum stationary Markov process
for the Schr\"odinger equation,
which corresponds to Brownian motion
for the diffusion equation, has not yet been established.
Formulating an analogous process to Brownian motion
in quantum mechanics would be useful for
estimating the time development of wave functions,
especially in higher degrees of freedom.

Because of the probabilistic nature of quantum mechanics, several stochastic formulations of quantum mechanics have been
investigated [6,7].
A typical example of such a stochastic process was
formulated by Nelson in 1966 [6].
In this method, the behavior of the constituent particles is determined
by a stochastic differential equation and their distribution
function coincides with the absolute square of the wave function.
In this sense, the trajectories in this method are those of
the observed particles.
We can apply this method to estimate the tunneling time
by making simulations of the trajectories [8].
However, the drift term in the stochastic equation is
calculated from the wave function,
and hence the Schr\"odinger equation must be solved to
construct the process.
A general method to construct a stochastic process
without solving the Schr\"odinger equation was
developed by Guerra and Marra [9].
In their method, the total distribution of particles is required to
determine the next step of a constituent particle,
though obtained process is a Markov process.
It is possible to determine the time evolution of all particles
simultaneously, but each individual
particle cannot be treated independently.
These methods treat the distribution function directly,
i.e. the absolute square of a wave function.

A way to avoid this difficulty is to search for
a stochastic process for the time evolution equations
which are equivalent to the Schr\"odinger equation.
One such equation is the quantum Liouville equation for
the Wigner function.
In this paper, we discuss a stationary quantum stochastic
process for the Wigner function in a lattice phase space.
A similar approach was considered by Cohendet et al. [10].
They constructed the Wigner function on a lattice
constituting of an odd number of lattice points, and found a background
quantum Markov process.

The Wigner function is defined by Wigner as

\begin{equation}
\rho(q,p)
={1\over\pi\hbar}\int_{-\infty}^\infty dr e^{2ipr/\hbar}\psi^*(q+r)\psi(q-r)
\end{equation}

\noindent
in phase space,
where \(\psi(q)\) is a wave function [11,12].
The Wigner function gives the density distribution in the
configuration space if it is integrated along the momentum direction:

\begin{equation}
\int_{-\infty}^{+\infty}\rho(q,p)dp = |\psi(q)|^2,
\end{equation}

\noindent
and gives the density distribution in the momentum space
if integrated along the configuration direction:

\begin{equation}
\int_{-\infty}^{+\infty}\rho(q,p)dq = |\tilde{\psi}(p)|^2.
\end{equation}

\noindent
This property is the so--called marginality
of the Wigner function.
In this sense, the Wigner function can be regarded as a kind
of {\it distribution function} in phase space and plays an
important role in quantum mechanics, especially in
the quantum--classical correspondence.
However, the value of the Wigner function can be negative,
so that it is not a distribution function itself.

Cohendet et al. regarded a lattice with odd \(N\) lattice points
as a cyclic group \(Z_N\) of order \(N\) and constructed a
Wigner function in the lattice phase space  \(Z_N\times Z_N\),
i.e. the direct product of \(Z_N\) and its dual \((\cong Z_N\)).
The Wigner function can also be negative on a finite lattice,
but the value is bounded.
They extended the phase space \(Z_N\times Z_N\) by a dichotomic
variable \(\sigma\) of value \(\pm1\) and considered a new
function on the extended space adding a constant
to the Wigner function.
The function is normalized and positive as the original function is bounded,
so it can be regarded as a distribution function of particles
in the new space.
They derived the time evolution equation for the distribution
function from the quantum Liouville equation, which determines
the time evolution of the Wigner function, and constructed the background
stochastic process of the particles using the Guerra--Marra method.
Although the obtained process has a Markov property,
the whole distribution function is required
to determine the evolution of a constituent particle.
The transition probability is time dependent and not stationary.

In this paper we introduce a stationary Markov process
that enables us to treat all particles independently,
by extending the phase space by a phase factor \(U(1)\)
on an odd lattice.
The process has branching and vanishing points and
a finite time interval can be maintained between the branchings.

\section{Stationary quantum Markov process for the Wigner function}

\subsection{Wigner function on a lattice}

Let \(N\) be an odd number.
We define \(N\times N\) phase and shift
matrices \(Q, P\) as

\begin{equation}
Q=(Q_{nm})\equiv(\omega^n\bar{\delta}_{n,m})
=\left(\begin{array}{cccccccc}
\omega^{{N\over2}-{1\over2}} & & & & & & & \\
   & \cdot & & & & & & \\
   & & & \omega & & & & \\
   & & & & 1 & & & \\
   & & & & & \omega^{-1} & & \\
   & & & & & & \cdot & \\
   & & & & & & & \omega^{-{N\over2}+{1\over2}} \\
\end{array}\right),
\end{equation}

\begin{equation}
P=(P_{nm})\equiv(\bar{\delta}_{n+1,m})
=\left(\begin{array}{ccccccccccc}
 0 &       &       &   &       &       & 1 \\
 1 & \cdot &       &   &       &       &   \\
   & \cdot & \cdot &   &       &       &   \\
   &       & 1     & 0 &       &       &   \\
   &       &       & 1 & \cdot &       &   \\
   &       &       &   & \cdot & \cdot &   \\
   &       &       &   &       & 1     & 0 \\
\end{array}\right),
\end{equation}

\noindent
where \(m,n\in Z_N\) and are labeled from \((N-1)/2\) to
\(-(N-1)/2\) with step -1,
\(\bar{\delta}\) is the
Kronecker delta function modulo \(N\),
\(\omega\) is a primitive \(N\)-th root of unity, e.g.
\(\exp(2\pi i/N)\).
The matrices \(Q, P\) 
correspond to the phase and shift operators

\begin{equation}
e^{ia\hat{q}}, \hspace{10mm}
e^{ia\hat{p}},
\end{equation}

\noindent
respectively, in continuous space--time,
where \(a\) is a lattice constant.
The Weyl matrices \(W(m,n)\) and
Fano matrices \(\Delta(m,n)\) are defined as

\begin{equation}
W(m,n)=\omega^{-2mn}Q^{2n}P^{-2m},
\end{equation}

\begin{equation}
\Delta(m,n)=W(m,n)T,
\end{equation}

\noindent
where \(T\) is a matrix given by

\begin{equation}
T=(T_{nm})\equiv(\bar{\delta}_{n+m,0})
=\left(\begin{array}{ccccccc}
   & & & & & & 1 \\
   & & & & & \cdot & \\
   & & & & \cdot & & \\
   & & & 1 & & & \\
   & & \cdot & & & & \\
   & \cdot & & & & & \\
 1 & & & & & & \\
\end{array}\right).
\end{equation}

\noindent
The Weyl matrices \(W(m,n)\) can also be written as

\begin{equation}
W(m,n)
=\omega^{2mn}P^{-2m}Q^{2n}
=P^{-m}Q^{2n}P^{-m}
=Q^{n}P^{-2m}Q^{n}.
\end{equation}

The \(N^2\) Weyl matrices and Fano matrices are both complete
and orthonormal in the trace norm:
\[
{\rm Tr}[W^\dagger(m,n)W(m,n)]=N\delta_{m,n},
\]
\[
{\rm Tr}[\Delta^\dagger(m,n)\Delta(m,n)]=N\delta_{m,n},
\]
\[
W^\dagger(m,n)=W(-m,-n),
\]
\begin{equation}
\Delta^\dagger(m,n)=\Delta(m,n),
\end{equation}
so that arbitrary matrices can be expressed as a linear combination
of these matrices.
We call the coefficients that appear in the Weyl and Fano
expansions the Weyl and Fano coefficients, respectively.

In the continuous case,
the Weyl operator is defined as

\[
W_c(q,p) = 2e^{-2ipq/\hbar}Q_c^{2p}P_c^{-2q}
= 2e^{2ipq/\hbar}P_c^{-2q}Q_c^{2p}
\]
\begin{equation}
=2Q_c^pP_c^{-2q}Q_c^p
=2P_c^{-q}Q_c^{2p}P_c^{-q}
\end{equation}

\noindent
where

\begin{equation}
Q_c=e^{i\hat{q}},
\hspace{10mm}
P_c=e^{i\hat{p}}.
\end{equation}

\noindent
The Wigner function is written as

\begin{equation}
\rho_c(q,p)={1\over2\pi\hbar}{\rm Tr}[\rho W_c(q,p)T_c]
\end{equation}

\noindent
where the operator \(T_c\) is given by

\begin{equation}
T_c\equiv\int_{-\infty}^\infty dq |-q\rangle\langle q|.
\end{equation}

\noindent
The Wigner function on a lattice is defined in an analogous way
to the continuous case:

\begin{equation}
\rho(m,n)
={1\over N}{\rm Tr}[\rho W(m,n)T]
={1\over N}{\rm Tr}[\rho \Delta(m,n)].
\end{equation}

\noindent
It is easily seen that this definition satisfies the marginality
condition on the lattice.

\subsection{Quantum Liouville equation for the Wigner function}

Let \(\tilde{H}(m,n)\) be the Weyl coefficients of the Hamiltonian,

\begin{equation}
H=\sum_{m,n}W(m,n)\tilde{H}(m,n),
\end{equation}

\noindent
and \(\rho(m,n)\) be the Fano coefficients of the density matrix,
i.e. the Winger function,

\begin{equation}
\rho=\sum_{m,n}\Delta(m,n)\rho(m,n).
\end{equation}

\noindent
The quantum Liouville equation

\begin{equation}
i\hbar{d\over dt}\rho=[H,\rho]
\end{equation}

\noindent
can be written using these coefficients as

\[
{d\over dt}\rho(m,n)={1\over i\hbar}\sum_{m',n'}
\{\tilde{H}(m-m',n-n')\omega^{-2(mn'-nm')}
\]
\begin{equation}
 -\tilde{H}^*(m-m',n-n')\omega^{2(mn'-nm')}\}
  \rho(m',n').
\end{equation}

We modify the equation as follows for later convenience.
We perform a polar decomposition of the Weyl coefficients
of the Hamiltonian:

\begin{equation}
\tilde{H}(m,n)=\tilde{h}(m,n)\omega^{\theta(m,n)},
\end{equation}

\noindent
where \(\theta(0,0)\) is 0 from the Hermiticity of the Hamiltonian.
The time evolution equation for the Wigner function
(20)
can be rewritten by this decomposition as follows:

\[
{d\over dt}\rho(m,n)={1\over \hbar}\sum_{m',n'}
\tilde{h}(m-m',n-n')
\]
\begin{equation}
\hspace{-10mm}
  \times\{\omega^{-2(mn'-nm')+\theta(m-m',n-n')-N/4}
   +\omega^{2(mn'-nm')-\theta(m-m',n-n')+N/4}\}
  \rho(m',n').
\label{eq:QLE-WG}
\end{equation}

\noindent
The term in the right-hand side equals zero when \(m=m', n=n'\),
so that it can be omitted in the sum.

\subsection{Generator of the Markov process and time evolution equation
for \(\varphi\)}

We extend the phase space \(Z_N\times Z_N\) with phase factor \(U(1)\)
and consider a non negative function \(\varphi(\alpha,m,n)\) on
\(U(1)\times Z_N\times Z_N\) that has the following relation with the Wigner
function:

\begin{equation}
\rho(m,n) = \int d\alpha e^{i\alpha} \varphi(\alpha,m,n).
\end{equation}

\noindent
Since the Wigner function is bounded in modulus by \(1/N\) 
on a lattice
\[-{1\over N}\le\rho(m,n)\le{1\over N},\]
the function \(\varphi(\alpha,m,n)\) exists,
although it is not uniquely determined.
We can take

\begin{equation}
\varphi(\alpha,m,n)=\varphi(-\alpha,m,n)
\end{equation}

\noindent
without loss of generality as \(\rho(m,n)\) is real.

We consider the following time evolution equation for \(\varphi(\alpha,m,n)\)
within a short time period \(\Delta t\):

\begin{equation}
\varphi(\alpha,m,n;t+\Delta t)
=
\int d\alpha'\sum_{m',n'}
M_{\Delta t}(\alpha,m,n;\alpha',m',n')\varphi(\alpha',m',n';t)
\end{equation}

\noindent
where \(M_{\Delta t}(\alpha,m,n;\alpha',m',n')\) is given by

\[
M_{\Delta t}(\alpha,m,n;\alpha',m',n') =
(1-D\Delta t)\delta(\alpha-f_{\Delta t}(\alpha'))\delta_{m,m'}\delta_{n,n'}
\]
\[
+{D\Delta t\over2|\tilde{h}|}\tilde{h}(m-m',n-n')
\]
\[
\times\{\delta(\alpha-\alpha'-2\pi(-2(mn'-nm')
+\theta(m-m',n-n')-N/4)/N)
\]
\begin{equation}
+\delta(\alpha-\alpha'-2\pi(+2(mn'-nm')
-\theta(m-m',n-n')+N/4)/N)\},
\end{equation}

\noindent
with

\begin{equation}
|\tilde{h}|\equiv \sum_{m,n}\tilde{h}(m,n).
\end{equation}

\noindent
We assume the function \(f_{\Delta t}(\alpha)\) in this equation satisfies

\begin{equation}
(1-D\Delta t)\cos f_{\Delta t}(\alpha)=\cos \alpha,
\end{equation}

\begin{equation}
\lim_{\Delta t \rightarrow 0} f_{\Delta t}(\alpha) = \alpha.
\end{equation}

The definition of \(M_{\Delta t}(\alpha,m,n;\alpha',m',n')\)
shows that particles with ratio \(D\Delta t\) move to other phase points
with probability \(\tilde{h}(m-m',n-n')/|h|\)
within \(\Delta t\).
On the other hand, other particles stay at the same phase
points and change their phase from \(\alpha\)
to \(f_{\Delta t}(\alpha)\) deterministically.
This means that \(M_{\Delta t}\) represents
the time evolution of a Markov process for \(\varphi(\alpha,m,n)\).

\subsection{Time evolution of the Wigner function}

We consider the time evolution of the Wigner function \(\rho(m,n)\)
when \(\varphi(\alpha,m,n)\) evolves following Eq. (25):

\[
\rho(m,n;t+\Delta t) =
\int d\alpha \int d\alpha' \sum_{m',n'}
e^{i\alpha}M_{\Delta t}(\alpha,m,n;\alpha',m',n') \varphi(\alpha',m',n';t)
\]
\[
= \rho(m,n;t)
  +{D\Delta t\over2|\tilde{h}|} \int d\alpha' \sum_{m',n'} \tilde{h}(m-m',n-n')
\]
\[
  \times(e^{i\alpha'}\omega^{-2(mn'-nm')+\theta(m-m',n-n')-N/4}
\]
\begin{equation}
  +e^{i\alpha'}\omega^{+2(mn'-nm')-\theta(m-m',n-n')+N/4})
  \varphi(\alpha',m',n';t).
\end{equation}

\noindent
We have

\[
{1\over\Delta t}\{\rho(m,n;t+\Delta t)-\rho(m,n;t)\}
\]
\[
= {D\over2|\tilde{h}|} \sum_{m',n'} \tilde{h}(m-m',n-n')
\]
\[
  \times(\omega^{-2(mn'-nm')+\theta(m-m',n-n')-N/4}
  +\omega^{+2(mn'-nm')-\theta(m-m',n-n')+N/4})
\]
\begin{equation}
  \times\int d\alpha' e^{i\alpha'} \varphi(\alpha',m',n';t)
\end{equation}

\noindent
and in the \(\Delta t\rightarrow 0\) limit

\[
{d\over dt}\rho(m,n)
= {D\over2|\tilde{h}|i}\sum_{m',n'} \tilde{h}(m-m',n-n')
\]
\begin{equation}
\hspace{-15mm}
  \times(\omega^{-2(mn'-nm')+\theta(m-m',n-n')-N/4}
  +\omega^{+2(mn'-nm')-\theta(m-m',n-n')+N/4})\rho(m',n').
\end{equation}

This shows that if \(\varphi(\alpha,m,n)\) evolves following
the time evolution equation
(25) and we take

\begin{equation}
D/2|\tilde{h}| = 1/\hbar,
\end{equation}

\noindent
the function \(\rho(m,n)\) obtained from the distribution function
\(\varphi(\alpha,m,n)\)
satisfies the quantum Liouville equation for the Wigner function.

\subsection{Generator of the Markov process for \(\varphi\)}

The generator of a Markov process \\
\(A(\alpha,m,n;\alpha',m',n')\) is generally defined by

\begin{equation}
A(\alpha,m,n;\alpha',m',n')\equiv
\lim_{\Delta t\rightarrow 0}{M_{\Delta t}-I \over \Delta t}.
\end{equation}

\noindent
The generator corresponding to \(M_{\Delta t}(\alpha,m,n;\alpha',m',n')\)
is given by

\[
A(\alpha,m,n;\alpha',m',n')=
{D\over\tan\alpha'}\delta'(\alpha-\alpha')\delta_{m,m'}\delta_{n,n'}
\]
\[
-D\delta(\alpha-\alpha')\delta_{m,m'}\delta_{n,n'}
+{D\over2|\tilde{h}|}\tilde{h}(m-m',n-n')
\]
\[
\hspace{-10mm}
\times\{\delta(\alpha-\alpha'-2\pi(-2(mn'-nm')
+\theta(m-m',n-n')-N/4)/N)
\]
\begin{equation}
\hspace{-10mm}
+\delta(\alpha-\alpha'-2\pi(+2(mn'-nm')
-\theta(m-m',n-n')+N/4)/N)\}.
\end{equation}

This generator is singular at \(\alpha=0, \pi\).
These singularities come from the fact that
when we set

\begin{equation}
f_{\Delta t}(\alpha) = \alpha + \Delta\alpha
\end{equation}

\noindent
in Eq. (28) and Eq. (29), we have

\begin{equation}
\Delta t/\Delta\alpha = -{1\over D}\tan\alpha
=0
\end{equation}

\noindent
and \(\Delta t\) becomes zero at these points.
This means we cannot take the time evolution for
any choice of \(\Delta\alpha\) at these points.
To avoid this difficulty, we have to choose the
initial \(\varphi(\alpha,m,n)\) to be zero at \(\alpha=0, \pi\)
and perform branching without changing the contribution
to the Wigner function when a particle reaches these points.
For example, a particle which reaches \(\alpha=0\) can
be replaced by two particles at \(\alpha=+\pi/3\) and
\(\alpha=-\pi/3\).

In the limit \(\Delta t\rightarrow 0\),
the time evolution equation for
\(\varphi(\alpha,m,n)\) satisfies

\[
\dot{\varphi}(\alpha,m,n)
=
{\partial\over\partial\alpha}
{D\over\tan\alpha}
\varphi(\alpha,m,n)
-D\varphi(\alpha,m,n)
\]
\[
+\int d\alpha'\sum_{m',n'}
{D\over2|\tilde{h}|}\tilde{h}(m-m',n-n')
\]
\[
\hspace{-10mm}
\times\{\delta(\alpha-\alpha'-2\pi(-2(mn'-nm')
+\theta(m-m',n-n')-N/4)/N)
\]
\[
\hspace{-10mm}
+\delta(\alpha-\alpha'-2\pi(+2(mn'-nm')
-\theta(m-m',n-n')+N/4)/N)\}
\]
\begin{equation}
\times\varphi(\alpha',m',n').
\end{equation}

The quantum Liouville equation for the Wigner function
\(\rho(m,n)\) also can be derived directly from this equation.
We observe that Eq. (38) reduces to Eq. (22)
if the part

\begin{equation}
{\partial\over\partial\alpha}
{D\over\tan\alpha}
\varphi(\alpha,m,n)
-D\varphi(\alpha,m,n)
\end{equation}

\noindent
vanishes after the integration with respect to \(\alpha\)
multiplying \(e^{i\alpha}\).
Paying attention to \(\varphi(\alpha)=\varphi(-\alpha)\),
the integration

\[
\int_0^{2\pi} e^{i\alpha} {\partial\over\partial\alpha} {1\over\tan\alpha}
\varphi(\alpha) d\alpha
\]

\noindent
becomes

\[
2\int_0^{\pi} \cos\alpha {\partial\over\partial\alpha} {1\over\tan\alpha}
\varphi(\alpha) d\alpha
=
2[\sin\alpha\varphi(\alpha)]_0^\pi
-2\int_0^{\pi} {\partial\cos\alpha\over\partial\alpha}{1\over\tan\alpha}\varphi(\alpha) d\alpha
\]
\begin{equation}
=
2\int_0^{\pi} {\cos\alpha} \varphi(\alpha) d\alpha
=
\int_0^{2\pi} e^{i\alpha} \varphi(\alpha) d\alpha
\end{equation}

\noindent
and Eq. (22) holds.

\section{Construction procedure}

In the following we summarize the procedure to determine
the time evolution of the Wigner function
from the Markov process of particles
with distribution function \(\varphi(\alpha,m,n)\).

\subsection{Preparation}

\noindent
{\bf 1-1)}
Prepare an additional space \(U(1)\) on each phase space
point \((m,n)\). \\

\noindent
{\bf 1-2)}
Calculate the Wigner function from the initial wave function. \\

\noindent
{\bf 1-3)}
Expand the Wigner function into the distribution function
and make an appropriate choice of \(\varphi(\alpha,m,n)\). \\

\noindent
{\bf 1-4)}
Perform the Weyl expansion of the Hamiltonian and calculate
the Weyl coefficients \(\tilde{H}(m,n)\). \\

\noindent
{\bf 1-5)}
Perform the polar decomposition of \(\tilde{H}(m,n)\)
and calculate \(\theta(m,n)\).

\subsection{Rules of the process}

\noindent
{\bf 2-1)}
Particles of ratio \(D\Delta t\) move to other phase points within \(\Delta t\).
Others stay at the same phase point. \\

\noindent
{\bf 2-2)}
The probability \(P(m,n;m',n')\)
for moving from \((m',n')\) to \((m,n)\) is determined by

\begin{equation}
P(m,n;m',n')=\tilde{h}(m-m',n-n')/|\tilde{h}|,
\end{equation}

\noindent
with Eq. (27). \\

\noindent
{\bf 2-3)}
The phase change is deterministic for particles without
jumps to other phase points.
The phase is rotated from \(\alpha\) to \(f_{\Delta t}(\alpha)\). \\

\subsection{Simulation}

\noindent
{\bf 3-1)}
Prepare an ensemble of particles with the initial
distribution function. \\

\noindent
{\bf 3-2)}
Make transitions for each particle independently in the
extended phase space following the probability rule
above within a period of time. \\

\noindent
{\bf 3-3)}
For a jump to another phase point,
add the phases \(\omega^{-2(mn'-nm')+\theta(m-m',n-n')-N/4}\) \\
and \(\omega^{2(mn'-nm')-\theta(m-m',n-n')+N/4}\) with probability 1/2. \\

\noindent
{\bf 3-4)}
Pull back the particle when it reaches the stopping point
\(\alpha=0,\pi\) with branching. \\

\noindent
{\bf 3-5)}
After performing the process for each particle,
integrate the phase part multiplying \(e^{i\alpha}\) 
to the density distribution function
\(\varphi(\alpha,m,n)\).
The resultant function is the Wigner function \(\rho(m,n)\)
under Eq. (33). \\

\noindent
The ensemble must be remade if the number of particles increases.

We show an example for \(N=3\) just as a clue to applications.
When the Hamiltonian and initial wave function are given by
\[
H=S_z=
\left(
\begin{array}{ccc}
1 & 0 & 0 \\
0 & 0 & 0 \\
0 & 0 & -1 \\
\end{array}
\right),
\hspace{10mm}
|\psi\rangle=
\frac{1}{2}\left(
\begin{array}{c}
1 \\
\sqrt{2} \\
1 \\
\end{array}
\right),
\]
with
\[
<S_x>=1, <S_y>=<S_z>=0,
\]
where
\[
S_x=
\frac{1}{\sqrt{2}}
\left(
\begin{array}{ccc}
0 & 1 & 0 \\
1 & 0 & 1 \\
0 & 1 & 0 \\
\end{array}
\right),
S_y=
\frac{i}{\sqrt{2}}
\left(
\begin{array}{ccc}
0 & -1 & 0 \\
1 & 0 & -1 \\
0 & 1 & 0 \\
\end{array}
\right),
\]
\noindent
the Wigner function and an extended distribution function
are given by
\[
\rho(m,n)=\frac{1-\sqrt{2}}{12},
\hspace{5mm}
\varphi(\alpha,m,n)=\frac{\sqrt{2}-1}{12}
  \left(\delta(\alpha-\frac{2\pi}{3})
       +\delta(\alpha+\frac{2\pi}{3})\right),
\]
for \((m,n)\)=(1,1),(1,-1),(-1,1),(-1,-1),
\[
\rho(m,n)=\frac{1+2\sqrt{2}}{12},
\hspace{5mm}
\varphi(\alpha,m,n)=\frac{1}{12}
  \left(\delta(\alpha-\frac{\pi}{3})
       +\delta(\alpha+\frac{\pi}{3})\right),
\]
for \((m,n)\)=(0,1),(0,-1),
\[
\rho(m,n)=1/12,
\hspace{5mm}
\varphi(\alpha,m,n)=\frac{1+2\sqrt{2}}{12}
  \left(\delta(\alpha-\frac{\pi}{3})
       +\delta(\alpha+\frac{\pi}{3})\right),
\]
for \((m,n)\)=(1,0),(-1,0),
\[
\rho(m,n)=1/3,
\hspace{5mm}
\varphi(\alpha,m,n)=\frac{1}{\sqrt{3}}
  \left(\delta(\alpha-\frac{\pi}{3})
       +\delta(\alpha+\frac{\pi}{3})\right),
\]
for \((m,n)\)=(0,0).
The Weyl coefficents have only two non-zero values, i.e.,
\[
\tilde{H}(0,1)=i/\sqrt{3},
\hspace{10mm}
\tilde{H}(0,-1)=-i/\sqrt{3},
\]
hence the stochastic process is simply determined from
\[
P(0,1)=P(0,-1)=1/2,
\]
\[
\theta(0,1)=\theta(0,-1)=3/2.
\]

\section{Summary}

If we consider a lattice constituting of \(N\) odd lattice
points as a cyclic group of order \(N\),
the corresponding phase space is \(Z_N\times Z_N\).
By extending the phase space by a phase factor \(U(1)\),
we can construct a stationary Markov process
in the space \(U(1)\times Z_N\times Z_N\).
This process has the property that
multiplying the phase factor to the density distribution
function of the process and integrating the phase part
gives a function which satisfies the quantum Liouville
equation for the Wigner function.

There exist branching points at the phase \(0, \pi\), but we can
maintain a finite time interval before the branching
by making an appropriate choice of the initial condition
and branching method.
There is no contribution from the distribution at the phase \(\pm\pi/2\).
The process does not need to be continued for particles reaching these points
as the stochastic rule is symmetric around these points and they give
no contribution at future times.
In this sense, these points can be treated as vanishing points.
The estimation of branching and vanishing ratio is a problem for future studies.

\end{document}